\documentclass[aps,prl,reprint,superscriptaddress,twocolumn,showpacs]{revtex4-2}
\usepackage{graphicx}
\graphicspath{{../}}
\usepackage[utf8]{inputenc}
\usepackage[english]{babel}
\usepackage{color}
\usepackage{soul}
\usepackage[section]{placeins}
\usepackage{tabularx}
\usepackage{array}
\usepackage{xcolor}
\usepackage{bm}
\usepackage{hyperref}
\hypersetup{
   colorlinks=true,     
   linkcolor=blue,      
   citecolor=blue,      
   filecolor=blue,   
   urlcolor=blue       
}

\usepackage{siunitx}
\DeclareSIUnit\oersted{Oe}

\newcommand{\eqref}[1]{(\ref{#1})}

\begin{document}

\title{Universal Critical Exponents of the Magnetic Domain Wall Depinning Transition}

\author{L. J. Albornoz}
\affiliation{Instituto de Nanociencia y Nanotecnolog\'{\i}a, CNEA--CONICET, Centro At\'omico Bariloche, Av. E. Bustillo 9500 (R8402AGP), San Carlos de Bariloche, R\'{\i}o Negro, Argentina.}
\affiliation{Universit\'e Paris-Saclay, CNRS,  Laboratoire de Physique des Solides, 91405, Orsay, France.}
\affiliation{Instituto Balseiro, Universidad Nacional de Cuyo--CNEA, Centro At\'omico Bariloche, Av. E. Bustillo 9500 (R8402AGP) San Carlos de Bariloche, R\'{\i}o Negro, Argentina.}
\author{E. E. Ferrero}
\affiliation{Instituto de Nanociencia y Nanotecnolog\'{\i}a, CNEA--CONICET, Centro At\'omico Bariloche, Av. E. Bustillo 9500 (R8402AGP), San Carlos de Bariloche, R\'{\i}o Negro, Argentina.}
\author{A. B. Kolton}
\affiliation{Instituto Balseiro, Universidad Nacional de Cuyo--CNEA, Centro At\'omico Bariloche, Av. E. Bustillo 9500 (R8402AGP) San Carlos de Bariloche, R\'{\i}o Negro, Argentina.}
\affiliation{Centro At\'omico Bariloche, Comisión Nacional de Energía Atómica (CNEA), Consejo Nacional de Investigaciones Cient\'{\i}ficas y T\'ecnicas (CONICET), Av. E. Bustillo 9500 (R8402AGP) San Carlos de Bariloche, R\'{\i}o Negro, Argentina.}
\author{V. Jeudy}
\affiliation{Universit\'e Paris-Saclay, CNRS,  Laboratoire de Physique des Solides, 91405, Orsay, France.}
\author{S. Bustingorry}
\affiliation{Instituto de Nanociencia y Nanotecnolog\'{\i}a, CNEA--CONICET, Centro At\'omico Bariloche, Av. E. Bustillo 9500 (R8402AGP), San Carlos de Bariloche, R\'{\i}o Negro, Argentina.}
\author{J. Curiale}
\email{curiale@cab.cnea.gov.ar, he/him/his}
\affiliation{Instituto de Nanociencia y Nanotecnolog\'{\i}a, CNEA--CONICET, Centro At\'omico Bariloche, Av. E. Bustillo 9500 (R8402AGP), San Carlos de Bariloche, R\'{\i}o Negro, Argentina.}
\affiliation{Instituto Balseiro, Universidad Nacional de Cuyo--CNEA, Centro At\'omico Bariloche, Av. E. Bustillo 9500 (R8402AGP) San Carlos de Bariloche, R\'{\i}o Negro, Argentina.}
\date{\today}
\begin{abstract}
Magnetic field driven domain wall dynamics in a ferrimagnetic GdFeCo thin film with perpendicular magnetic anisotropy is studied using low temperature magneto-optical Kerr microscopy. 
Measurements performed in a practically athermal condition allow for the direct experimental determination of the velocity ($\beta = 0.30 \pm 0.03$) and correlation length ($\nu  = 1.3 \pm 0.3$) exponents of the depinning transition. The whole family of exponents characterizing the transition is deduced, providing evidence that the depinning of magnetic domain walls is better described by the quenched Edwards-Wilkinson universality class.
\end{abstract}

\maketitle

The depinning of interfaces interacting with a weak pinning disorder is encountered in a large variety of physical systems (contact lines in wetting, crack fronts, ferromagnetic and ferroelectric domain walls, etc.)~\cite{FisherPhysRep1998, GiamarchiCRP2013, ChauvePRB2000, PonsonPRL2009, LeDoussal2009, LePriolPRL2020, wiese2021theory}. Understanding the underlying physics on how those interfaces get stuck, detach and move is critical. 
For instance, storage and spintronic devices based on the control of magnetic domain walls (DWs) are among the archetypal examples of applications based on the stability and precise positioning
of interfaces~\cite{HirohataReviewSpintronicJMMM2020, LuoNat2020, PueblaNatCommMat2020}.
Years of theoretical and experimental studies on the field of depinning have taken advantage of simple statistical models that capture such a universally observed behavior.
Yet, it still remains highly non-trivial to embrace the idea that phenomena involving dramatically different scales and microscopic physical interactions can actually be quantitatively described by those simple models.

Magnetic DWs moving in thin ferromagnetic films with perpendicular anisotropy, constitute a well known model system to test theoretical predictions, given that they essentially behave as weakly randomly pinned elastic lines in a 2-dimensional medium.
Since the seminal work of Lemerle {\it et al.}~\cite{LemerlePRL1998}, most of the experimental studies on driven DWs have dealt with the thermally activated creep regime, where the velocity follows an Arrhenius law depending on temperature $T$ and driving force $f$, $\ln v \sim 1 /(T f^\mu)$; $\mu$ is an universal exponent.
Successfully, the measured values of $\mu$ reported in the literature for model experimental systems~\cite{LemerlePRL1998, KimNat2009, GorchonPRL2014, ChoiSciRep2016, DiazPardoPRB2017, GrassiPRB2018, ShahbaziPRB2019} are in agreement with the theoretical predictions~\cite{ChauvePRB2000, FerreroCRP2013}. 
However, the picture stands incomplete.
The {\it equilibrium} universality class (that determines $\mu$) is identically shared by two different Hamiltonian models for the description of driven elastic interfaces.
Based on a competition between elasticity and quenched disorder, both models assume short-ranged elastic and pinning forces, but elasticity is harmonic in one case and anharmonic in the other, leading to different depinnning peculiarities. 
Therefore, the agreement between predicted and observed creep 
regimes still does not tell us whether for depinning we should expect
the critical exponents corresponding to either of the two strongest candidates, the harmonic elasticity, viz., the quenched Edwards-Wilkinson (qEW) {\it depinning} universality class~\cite{LeDoussal2002}, or the one resulting from the anharmonic elasticity, viz., the quenched Kardar-Parisi-Zhang (qKPZ) {\it depinning} universality class~\cite{RossoPRL2001,RossoPRE2003}.
In order to discern which model better describes not only the creep
regime but also the {\it depinning regime} observed in experiments, 
we hereafter tackle the direct measurement of depinning exponents.

The depinning regime is best manifested at zero temperature: creep is suppressed and a 
depinning threshold force $ f_d $ separates a zero velocity ($v=0$) state for $ f<f_d $, from a finite velocity ($v>0$) state 
for $ f>f_d $~\cite{BuldyrevPRA1992,TangPRA1992,LeDoussal2002,RossoPRE2003,LeeJKPS2005,DuemmerPRE2005,DuemmerJStatMech2007,FerreroPRE2013,FerreroPRL2017,wiese2021theory}.
Above the threshold, the velocity is expected to vary as a power law with the force $v \sim (f-f_d)^\beta$, with $\beta$ being the velocity critical exponent.
Concomitantly, when $f_d$ is approached from above, a correlation length characterizing the transition diverges as $(f-f_d)^{-\nu}$ where $\nu$ is known as the correlation length critical exponent.
Recently, universal depinning behavior has been evidenced explicitly for a wide set of materials~\cite{DiazPardoPRB2017}. 
However, the determination of critical exponents as $\beta$ or $\nu$ has been restricted due to finite temperature effects. 
Basically, thermal activation produces a rounding of the depinning transition which impedes their direct determination. 
 
In this Letter, we study the DW dynamics in a ferrimagnetic thin film, with the experimental ability to decrease 
the temperature to values such that the velocity-force characteristics is close to the ideal athermal case.
We experimentally measure $\beta$ and $\nu$, which allow us to determine the whole set of critical exponents 
of the depinning transition and allocate the studied model system in the qEW universality class.

\textit{Experimental details -}
The studied sample consists in a Ta(5~nm)/Gd$_{32}$Fe$_{61.2}$Co$_{6.8}$(10~nm)/Pt(5~nm) trilayer deposited on a thermally oxidized silicon SiO$_2$(100~nm) substrate by RF sputtering. 
The GdFeCo thin film is a rare earth-transition metal (RE-TM) ferrimagnetic compound presenting perpendicular magnetic anisotropy with the RE and TM magnetic moments antiferromagnetically coupled.
DW motion is measured using Polar Magneto-Optical Kerr Effect (PMOKE) microscopy. 
We obtain images of DWs before and after the application of square-shaped magnetic field pulses of amplitude $ H $ and duration $ \Delta t $, and measure the resulting mean DW displacement normal to the DW profile $\Delta x_n$.
The mean DW velocity $ v $, characterizing its collective motion, at a given field $ H $ is calculated as $\Delta x_n/\Delta t$.
The maximum investigated field is limited by the nucleation of multiple magnetic domains, which impedes clear DW displacement measurements.
Additionally, the microscope is equipped with a cryostat and a temperature controller that give us access to a wide temperature range, from $ 4~\si{\kelvin} $ to $ 360~\si{\kelvin} $.

\begin{figure}[t!]
\includegraphics[width=0.85\columnwidth]{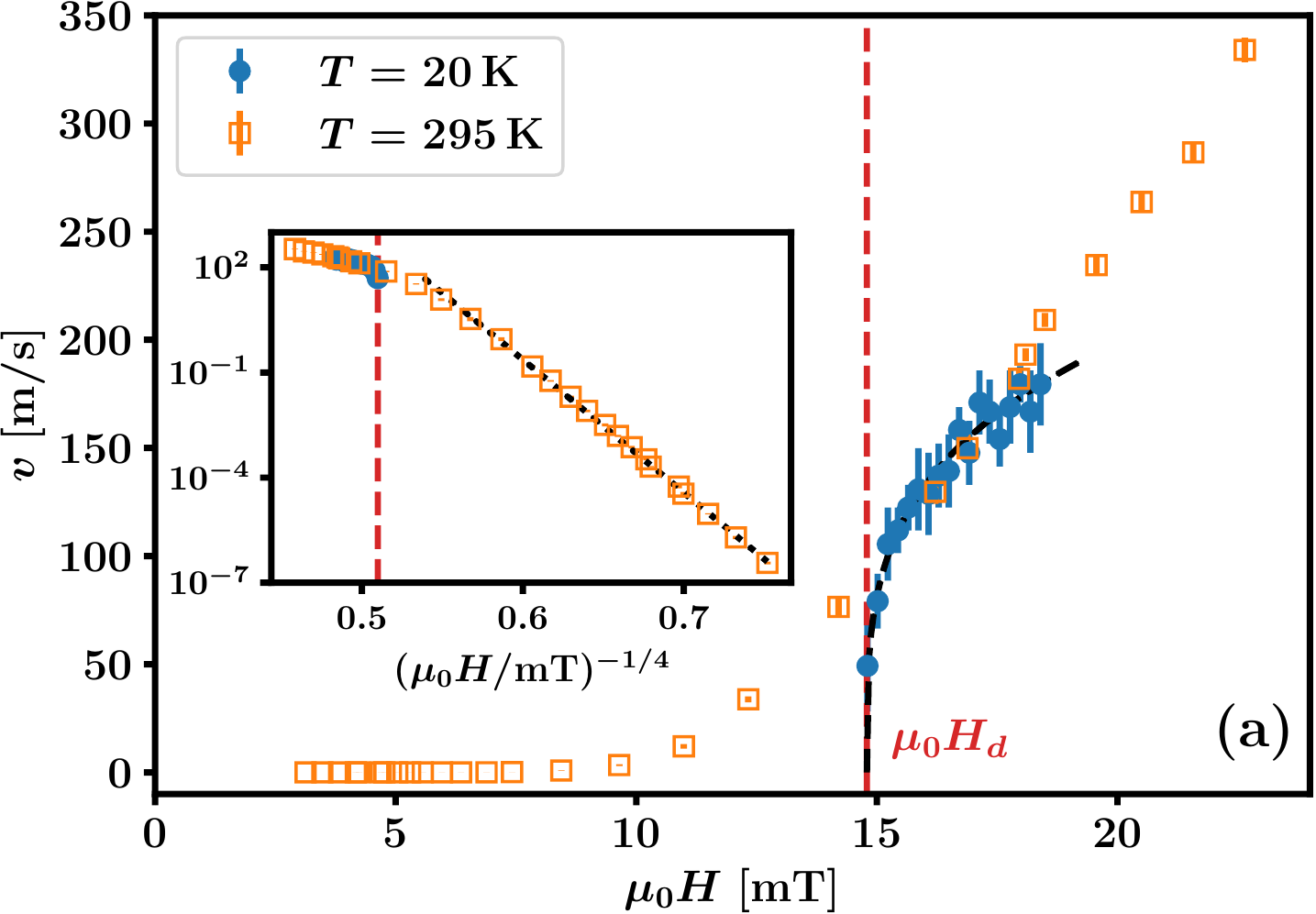}
\includegraphics[width=0.85\columnwidth]{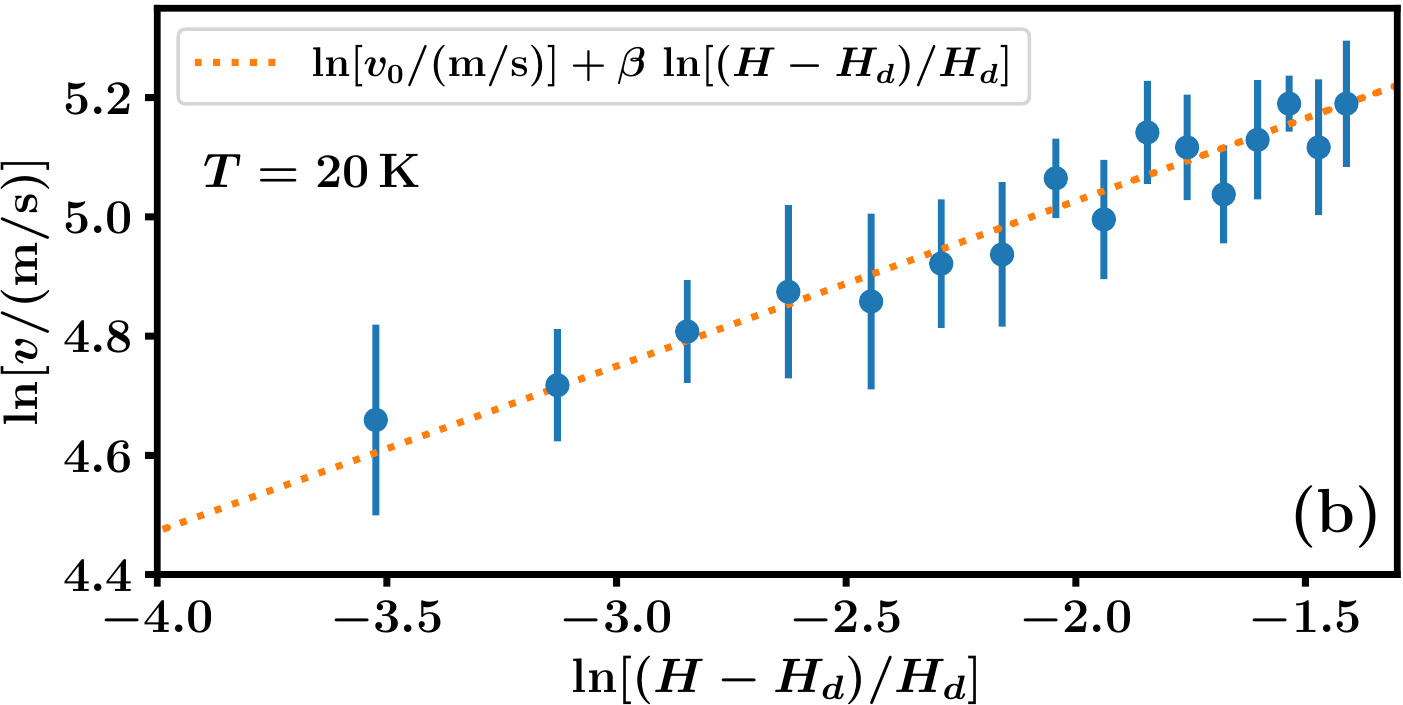}
\caption{\label{fig:VdwvsH_20Kand295K}
High- and low-temperature velocity curves. (a) Domain wall velocity ($ v $) as a function of the external magnetic field ($ \mu_0 H $) for a GdFeCo thin film at $ T = 20~\si{\kelvin} $ (circles) and $ T = 295~\si{\kelvin} $ (squares).
The depinning field $ \mu_0 H_d $ corresponding to $ T = 20~\si{\kelvin} $ (vertical dashed red line), and the fit of Eq. \eqref{eq:DepinningVelocity} for that temperature (dashed black line) are indicated. 
In the inset, in semi-logarithmic scale, the thermally activated creep regime observed for $ T = 295~\si{\kelvin} $ is emphasized with the dotted line, a fit to the creep formula.
(b) Results obtained at $ T = 20~\si{\kelvin} $ and plotted in a logarithmic scale to illustrate an excellent agreement with Eq. \eqref{eq:DepinningVelocity}. The fitted parameters are $\beta = 0.28 \pm 0.08$, $\mu_0 H_d = 14.8 \pm 0.2~\si{\milli\tesla}$ and $v_0 = 270 \pm 40~\si{\meter/\second}$. 
}
\end{figure}

\textit{Zero-temperature-like depinning transition -}
The contribution of thermal effects on the depinning transition is shown in Fig.~\ref{fig:VdwvsH_20Kand295K}(a), which compares two velocity-field curves with very similar depinning threshold ($ H_d \approx 15 ~\si{\milli\tesla} $) measured at temperatures that differ in more than one order of magnitude ($ 20~\si{\kelvin} $ and $ 295~\si{\kelvin} $).
As it can be observed, thermal activation blurs the depinning transition.
Above $H_d$, velocities measured at both temperatures are rather similar. However, close to $H_d$ the velocity sharply decreases for $T= 20~\si{\kelvin}$ while it remains finite, even well below $H_d$, for $ T = 295~\si{\kelvin} $.~\footnote{In this case, applied magnetic fields smaller than the one producing the lowest reported velocity, yield ill-defined mean DW velocities. Some regions of the DW remain pinned during the observation window. Therefore, those points are not taken into account.}. 
The curve obtained for $ T = 295~\si{\kelvin} $ (Fig.~\ref{fig:VdwvsH_20Kand295K}(a)) presents both the typical 
thermally activated behavior for low fields and the thermal rounding around $H_d$, as has been
reported in the literature  for most of the field- and current-driven experiments~\cite{JeudyPRL2016, Diaz_Pardo_PRB_2019}.
Below the depinning threshold ($3 ~\si{\milli\tesla}<H< H_d$), the DWs follow a thermally activated creep dynamics. 
The velocity obeys the Arrhenius law $v(H,T) = v(H_d,T) e^{-\Delta E/k_B T}$, with $k_B$ the Boltzmann constant. 
The universal creep barrier is given by $\Delta E=k_B T_d \left[ \left({H}/{H_d} \right)^{-\mu}-1 \right]$~\cite{JeudyPRL2016}, 
where $T_d$ is the so-called \textit{depinning temperature} and $\mu$ is the creep exponent ($\mu=1/4$)~\cite{LemerlePRL1998, ChauvePRB2000, FerreroCRP2013}.
As observed in Fig.~\ref{fig:VdwvsH_20Kand295K}(a) (see the inset), the experimental data measured at 
$T = 295~\si{\kelvin}$, which covers more than eight orders of magnitude, are in good agreement with the creep law. 
Remarkably, the depinning temperature estimated from the fit, $T_d \approx 10000~\si{\kelvin}$, is among the largest $T_d$ reported in the literature~\cite{JeudyPRBTabla2018}. 
Moreover, for the lowest temperature at which we can observe a clean creep dynamics, $T = 100~\si{\kelvin}$,  
the value of $T_d$ is close to $\approx 30000~\si{\kelvin}$. 
This corresponds to a ratio $T_d/T$ ($>300$), which had never been reached experimentally~\cite{JeudyPRBTabla2018} so far. 

\textit{The universal velocity depinning exponent $\beta$ -}
The ratio $T_d/T$ is strongly enhanced as the temperature decreases, thus explaining the absence of creep regime 
and the quasi-athermal depinning behavior observed for the Ta/GdFeCo/Pt film at $T = 20~\si{\kelvin}$.
Then the velocity-field data at $T = 20~\si{\kelvin}$ can be analyzed as a zero temperature-like depinning transition.
Quantitative information is thus obtained by fitting the data using~\cite{BustingorryEPL2008, BustingorryPRB2012, FerreroCRP2013, PurrelloPRE2017, DiazPardoPRB2017, FerreroReview2021, SuppMat}
\begin{equation}
\label{eq:DepinningVelocity}
  v(H, T =0~\si{\kelvin})= v_0{\left(\frac{H-H_d}{H_d}\right)}^\beta.
\end{equation}
In Eq. \eqref{eq:DepinningVelocity}, the depinning velocity $ v_0 $ and $H_d$ are material and temperature dependent parameters~\cite{BustingorryPRB2012,DiazPardoPRB2017}.
As shown in Fig.~\ref{fig:VdwvsH_20Kand295K}(b), the data presents a good quantitative agreement with the prediction over the whole 
range of measured magnetic fields (see ~\cite{SuppMat} for details on the fitting procedure). 
Our direct experimental measurement of $\beta$ ($= 0.28 \pm 0.08$) somehow justifies 
the values assumed in former experimental studies of depinning~\cite{DiazPardoPRB2017} ($\beta= 0.25$),
and, more importantly, matches the exponent values predicted theoretically ($\beta \simeq 0.245 - 0.33$)~\cite{DuemmerPRE2005,FerreroPRE2013}. 

\begin{figure}[t!]
\includegraphics[width=0.9\columnwidth]{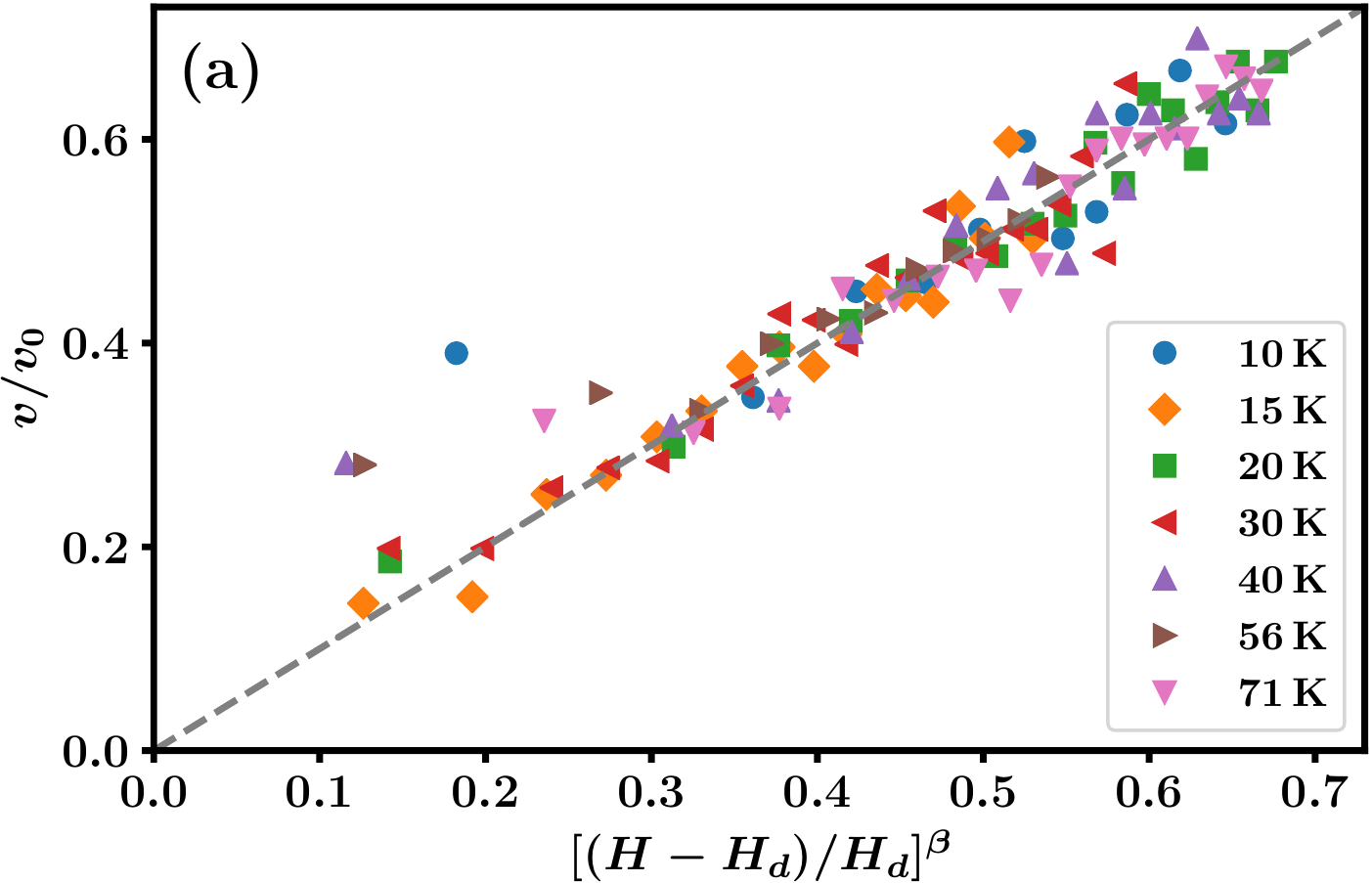}
\includegraphics[width=0.9\columnwidth]{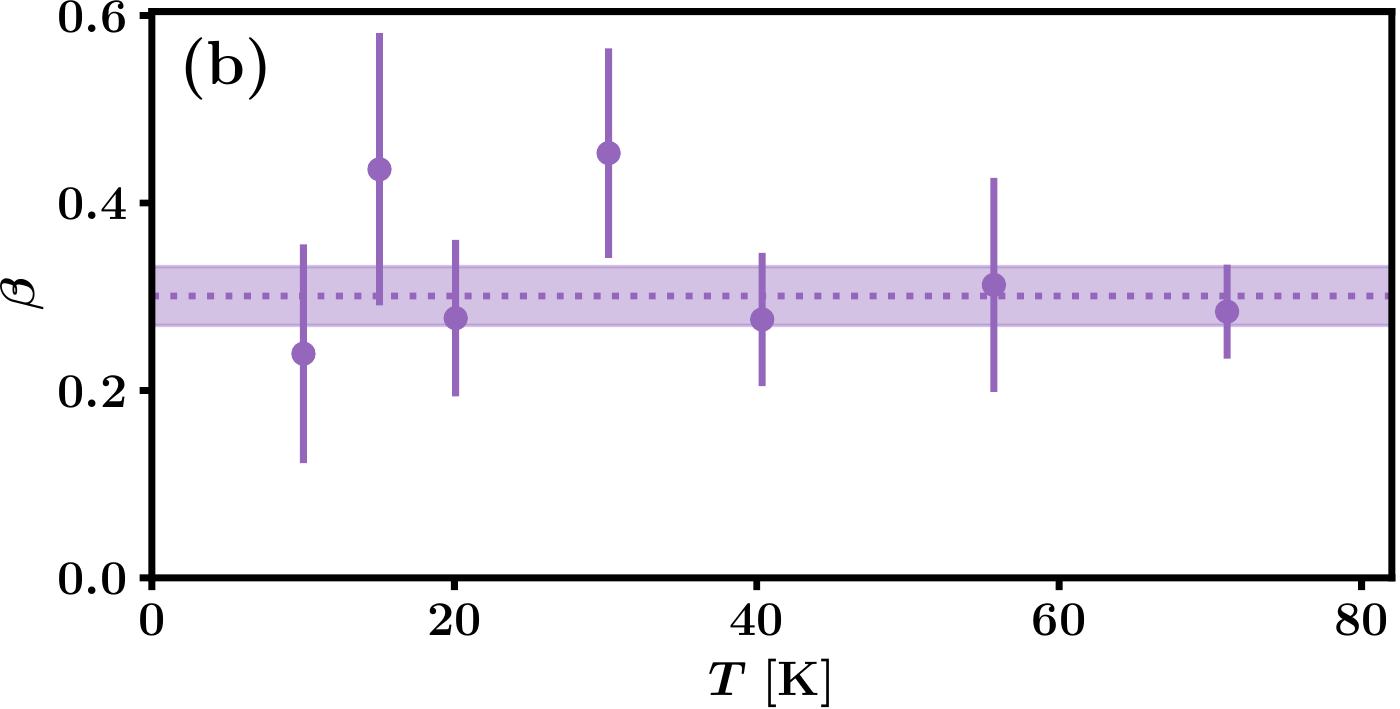}
\caption{\label{fig:BetavsTandRescaledVDWvsH}
	(a) Consistency with zero temperature depinning. 
  Rescaled velocity $v/v_0$ as a function of $ [(H-H_d)/H_d]^{\beta}$ in the temperature range [$10~\si{\kelvin}\leq$ T $\leq 71~\si{\kelvin}$], using the experimentally determined values $H_d$, $v_0$ and $\beta$. 
	(b) Velocity depinning exponent as a function of temperature. The mean value ($\overline{\beta} = 0.30 \pm 0.03$) is indicated by a dotted line and its uncertainty as a shaded area.
}
\end{figure}

In order to further increase and test the accuracy of the $\beta$ estimation, we have explored data on a broad 
temperature range in which the athermal depinning behavior is observed ($10~\si{\kelvin}\leq$ T $\leq 71~\si{\kelvin}$).
Following the analysis presented for $T=20~\si{\kelvin}$, the velocity curves obtained at different temperatures have been 
fitted independently (see ~\cite{SuppMat} for details on the method and the values of $v_0$ and $H_d$). 
Fig.~\ref{fig:BetavsTandRescaledVDWvsH}(a) shows a comparison of the velocity curves with rescaled axis, 
$v/v_0$ versus $[(H-H_d)/H_d]^\beta$. 
The data collapses onto a single master curve, which attests the common behavior observed for different 
$T$ and their good agreement with Eq.~\eqref{eq:DepinningVelocity}. 
Moreover, the obtained values of $\beta$ shown in Fig.~\ref{fig:BetavsTandRescaledVDWvsH}(b) do not present 
any particular trend with temperature but rather appear to fluctuate around its weighted mean value $\overline{\beta} = 0.30 \pm 0.03$.
For a verification of this result, we have performed an alternative fitting procedure consisting on a global 
fit of all the velocity-field points with $\beta$ as a ($T$-independent) free parameter, 
obtaining $\beta = 0.33 \pm 0.04$ (see ~\cite{SuppMat}). 
Both analysis protocols lead to very similar results also for the material dependent parameters $v_0$ and $H_d$. 
This allows to confirm that the measured value of $\beta$ for Ta/GdFeCo/Pt is temperature independent.

As a critical exponent is expected to be universal, it is particularly important to test the compatibility of the measured value of $\beta$ with other magnetic materials. 
For this purpose, a new look to the velocity curves obtained for ultrathin films made of Au/Co/Au~\cite{KirilyukJMMM1997} 
and (Ta/CoFeB/MgO)~\cite{BurrowesAPL2013} stands as an unavoidable task, due to the materials' large $T_d/T$-ratio ($=90-170$) and large magnetic field range of observation of the depinning, respectively~\cite{DiazPardoPRB2017}. 
A global fit as the one described above was performed to the data reported in the literature 
for both materials and the fitted values of $\beta$ (see ~\cite{SuppMat}) confirm the universal 
nature of the exponent. 

\begin{figure}[t!]
	\includegraphics[width=1\columnwidth]{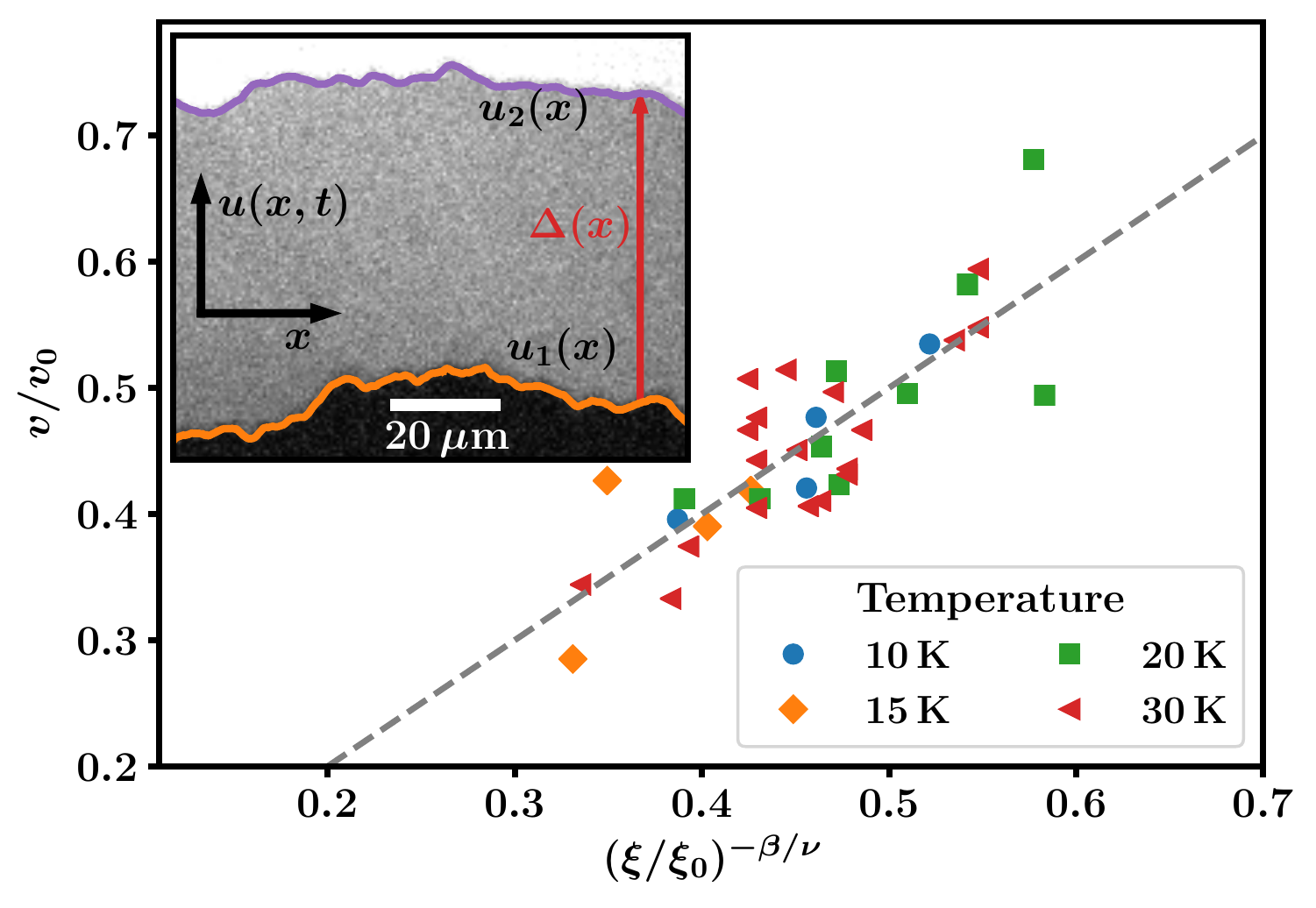}
	\caption{\label{fig:VdwvsCorrLength}
	Velocity/correlation-length exponent. $v/v_0$ vs. ($\xi/\xi_0)^{-\beta/\nu}$ for different external magnetic fields and temperatures $T = 10, 15, 20$ and $30~\si{\kelvin}$. Starting from the highest magnetic field used in this experiment, the correlation length	increases and the velocity decreases when the magnetic field approaches $H_d$. The relation between $v$ and $\xi$ can be described using a power-law, as indicated by the dashed line.	The mean value of the $\nu$ exponent in the studied temperature range is $\overline{\nu} = 1.3 \pm 0.3$. The inset shows two successive domain wall profiles used to compute the velocity correlation function at T = 20 K and $\mu_0 H = 15.44~\si{\milli\tesla}$. The magnetic field pulse duration is $\Delta t = 0.5~\si{\micro \second}$ and the corresponding velocity $v(x) = \Delta(x)/\Delta t = 112.3 \pm 0.3~\si{\meter/\second}$.
}
\end{figure}

\textit{Correlation length depinning exponent -} 
In order to complete the analysis of the universality of DW depinning, 
we also address the measurement of the exponent $\nu$ characterizing the correlation 
length divergence when $H \rightarrow H_d(T)$:
\begin{equation}
 \label{eq:DepinningCorrLength}
\xi(H,T = 0~\si{\kelvin}) = \xi_0 \left( \frac{H- H_d}{H_d} \right)^{-\nu},
\end{equation}
with $\xi_0$ a characteristic length.
At the depinning threshold the typical size of avalanches diverges~\cite{FerreroCRP2013} and, if two points on the DW belong to the same avalanche, their instantaneous velocities will be correlated~\cite{LeDoussal2013}.
Therefore, the typical length of avalanches shows up as the characteristic length scale in the velocity correlation function, and a divergence in the correlation length $\xi$ is a signature of diverging avalanche sizes.
To compute the velocity correlation function we consider consecutive DW profiles $u_1(x)$ and $u_2(x)$ at different temperatures. Both profiles are obtained after tilting the PMOKE images in order to orientate DWs in the horizontal direction, normal to the y-axis.
As shown in the inset of Fig.~\ref{fig:VdwvsCorrLength}, if $u_1(x)$ and $u_2(x)$ are the DW profiles before and after the application of a magnetic field pulse of intensity $H$ and duration $\Delta t = 0.5~\si{\micro \second}$, then $\Delta(x) = u_2(x) - u_1(x)$ is the displacement field.
Using the local velocity $v(x) = \Delta(x)/\Delta t$, the velocity correlation function, for profiles of longitudinal 
size $L=N \delta$ and pixel size $\delta$, is computed as~\cite{DuemmerPRE2005}
\begin{equation}
 C_{v}(x) = \Delta t^{-2} \sum_{x^{\prime}=0}^{\textcolor[rgb]{0.2,0,1.0}{{L-x}}} \left[\Delta\left(x^{\prime}+x\right)-\bar{\Delta}\right]\left[\Delta\left(x^{\prime}\right)-\bar{\Delta}\right],
\end{equation}
where $\bar\Delta = (1/N)\sum_x \Delta(x)$. 
The correlation length for different magnetic fields can be estimated from the definition $C_v(x = \xi) \approx C_v(x = 0)/2$~\cite{SuppMat}. 
Using the same two DW profiles, the field-dependent mean velocity can be estimated as $v = \bar\Delta/\Delta t$.
Now that we have access to $\xi(H)$ and $v(H)$, and combining Eqs.~\eqref{eq:DepinningVelocity} and \eqref{eq:DepinningCorrLength}, we can use the expression  
\begin{equation}
v = v_0 \left(\frac{\xi}{\xi_0}\right)^{-\beta/\nu}
\label{eq:BetaNu}
\end{equation}
to extract $\beta/\nu$ and $\xi_0$ for different temperatures.

Main panel of Fig.~\ref{fig:VdwvsCorrLength} shows $v/v_0$ vs. $(\xi/\xi_0)^{-\beta/\nu}$ for different external magnetic fields and temperatures. 
A power law behavior following Eq. \eqref{eq:BetaNu} is observed over a temperature range $10~\si{\kelvin} \leq T \leq 30~\si{\kelvin}$,
which is narrower than the range showing an effectively athermal reduced velocity (Fig. \ref{fig:BetavsTandRescaledVDWvsH}). 
This suggests, as a first observation, that the reduced correlation length is more sensitive to finite temperature effects than the reduced velocity.
Anyhow, fitting the data on Fig.~\ref{fig:VdwvsCorrLength} with power-laws we obtain the exponent $\beta/\nu$ and $\xi_0$ for different temperatures (see \cite{SuppMat} for details).
Noteworthy, in the studied temperature range, the mean value $\overline{\xi_0} = 90\pm 50~\si{nm}$ results of the same order of magnitude than the expected Larkin length: 
$L_c=\frac{k_B T_d}{2 M_s H_{d} s \xi_p}\approx 165~\si{nm}$~\footnote{Notice that even if $\xi_0$ is below the optical resolution ($\approx 1~\si{\micro\meter}$), this does not represent an inconsistency, since this quantity is a fitting parameter. The measured quantity $\xi$ is typically of several micrometers. Computing $\overline{\xi_0}$ allows us to check the theory self-consistency when comparing it with $L_c$.}, where $M_s$ is the saturation magnetization, $s$ the sample thickness and $\xi_p$ is the estimated correlation-length of the pinning-force in the direction of DW displacements~\cite{SuppMat}.
Hence the whole fitted velocity range satisfies $\xi \gg \xi_0 \sim L_c$, as theoretically expected for the critical regime at depinning~\cite{ChauvePRB2000}.
Finally, combining the independently extracted values for $\beta$ and $\beta/\nu$ we obtain, at different temperatures, the correlation length exponent $\nu$, whose average value results in $\overline{\nu} = 1.3 \pm 0.3$.

\textit{Universality class of DW depinning -}
Having determined independently two critical exponents, the full set of critical exponents (i.e., the universality family or universality class) for the depinning transition, can be deduced making use of scaling laws. Let us now compare the values of the exponents estimated from our measurements with those defining  qEW and qKPZ; the two strongest candidates for the universality class of drivenDWs in disordered media \cite{FerreroReview2021}. First notice that the experimentally estimated values ($\beta\simeq 0.30$, $\nu\simeq 1.30$) are significantly smaller
than those predicted for the quenched Kardar-Parisi-Zhang (qKPZ) universality class ($\beta \simeq 0.64$ and $\nu \simeq 1.73$,
as numerically obtained using either the directed percolation model~\cite{TangPRA1992, BuldyrevPRA1992}, 
anharmonic elasticity~\cite{RossoPRE2003}, or direct integration of the qKPZ equation~\cite{LeeJKPS2005}).
Furthermore, if linear elasticity were considered to be long-range, exponents $\beta \simeq 0.625$ and $\nu \simeq 1.625$~\cite{DuemmerJStatMech2007} would have been expected. 
Therefore, neither the qKPZ universality class nor the long-range elasticity quantitatively describe the physics of field driven DW depinning.
In contrast, the experimental values of $\beta$ and $\nu$ present a much better agreement with the qEW universality class (see Table~\ref{tab.Exponents}), with a model where DW elasticity is assumed to be short-range. 

\begin{table}[b!]
\caption{Summary and comparison of critical exponents
\label{tab.Exponents}}
 \begin{tabular}{cccl}
\hline
\hline
{Exp.} & {This work} & {qEW} & {qKPZ}\\
\hline
$\beta$ & $0.30 \pm 0.03$ & $0.245\pm0.006$\hfill\cite{FerreroPRE2013} & $\simeq 0.64$\hfill\cite{TangPRL1995, RossoPRE2003,FerreroReview2021}\\
&  & \hfill$0.33\pm0.02$\hfill\cite{DuemmerPRE2005}& \\
$\nu$ & $1.3 \pm 0.3$ & $1.333 \pm 0.007$\hfill\cite{FerreroPRE2013} & $\simeq 1.73$\hfill\cite{TangPRL1995}\\
$\zeta$ & $1.2 \pm 0.2$ & $1.250 \pm 0.005$\hfill\cite{FerreroPRE2013} &$\simeq 0.63$\hfill\cite{RossoPRE2003}\\
$z$ & $1.5 \pm 0.2$ &  $1.433 \pm 0.007$\hfill\cite{FerreroPRE2013} & \quad 1\hfill\cite{TangPRL1995}\\ 
$\tau$ & $1.11 \pm 0.07$ &  \hfill$1.11 \pm 0.04$\hfill\cite{FerreroPRL2017} &$\simeq 1.26$\hfill\cite{FerreroReview2021,TangPRL1995,RossoPRE2003}\\ 
\hline 
\hline 
 \end{tabular}
\end{table}
The independent determination of two depinning critical exponents (in this case $\beta$ and $\nu$ corresponding to the qEW universality class) permits to calculate, through scaling relations, the rest of the exponents of the family.  Using the scaling relations~\cite{ChauvePRB2000, FisherPhysRep1998, FerreroPRE2013} $\nu = 1/(2 - \zeta)$, $\beta = \nu (z - \zeta)$, and  $\tau = 2-2/(1 + \zeta)$, we determine the dynamical exponent $z$, the roughness exponent $\zeta$, and the avalanche-size distribution exponent $\tau$. These are reported in Table~\ref{tab.Exponents} for completeness, together with the measured values of $\beta$ and $\nu$.

{\it Discussion-}
The theoretical-computational qEW values reported in the literature for $\beta$ are themselves rather disperse.
The asymptotic value ($\beta=0.245 \pm 0.006$)~\cite{FerreroPRE2013} is expected to be reached only for a driving field sufficiently close to the depinning field, corresponding to correlation lengths much larger than the collective pinning length ($\xi \gg L_c$). 
Otherwise, scaling corrections are evidenced and a larger effective $\beta$ is expected~\cite{FerreroPRE2013}. 
In our case, the maximum measured value of $\xi$ ($\approx 30~\si{\micro\meter}$) is not much larger than the Larkin length ($\approx 165~\si{\nano\meter}$).
The measured value ($\beta= 0.30 \pm 0.03$) is slightly bigger than the asymptotic theoretical value, but close to qEW numerical simulations with a similar $\xi/L_c \sim 100$ ratio ($\beta=0.33 \pm 0.01$)~\cite{DuemmerPRE2005}.

In conclusion, we have shown that using a ferrimagnet with a very high depinning temperature, we were able to strongly reduce thermally activated effects around the depinning transition.
It granted us access not only to a direct experimental determination of the exponent associated to the order parameter of the depinning transition $\beta$ but also to the correlation length critical exponent $\nu$, and hence to the whole set of critical exponents. 
From our results it follows that the depinning transition of domain walls dynamics in a perpendicularly magnetized thin film  is better described by the universality class of the quenched Edwards-Wilkinson model with short range elasticity.

\begin{acknowledgments}
The authors would like to thank J. Gorchon, C. H. A. Lambert, S. Salahuddin and J. Bokor for kindly facilitating the high-quality samples used in this work. We also want to thank Kay Wiese for enlightening discussions and for reading the manuscript.
This work was partly supported by the Argentinian Project No. PICT2016- 0069 (MinCyT), PICT 2017-1202 and PICT 2017-0906 (MinCyT) and by the UNCuyo 2019 Grants No. 06/C561, No. 06/C578 and M083. 
\end{acknowledgments}

%

\end{document}